\documentclass{moriond}

\def\Journal#1#2#3#4{{#1} {\bf #2}, #3 (#4)}

\def\PRL{\em Phys. Rev. Lett.}
\def\PRD{{\em Phys. Rev.} D}

\def\be{\begin{equation}}
\def\ee{\end{equation}}
\def\bea{\begin{eqnarray}}
\def\eea{\end{eqnarray}}

\usepackage{graphicx,graphbox}

\newcommand{\Photo}{\includegraphics[height=35mm]{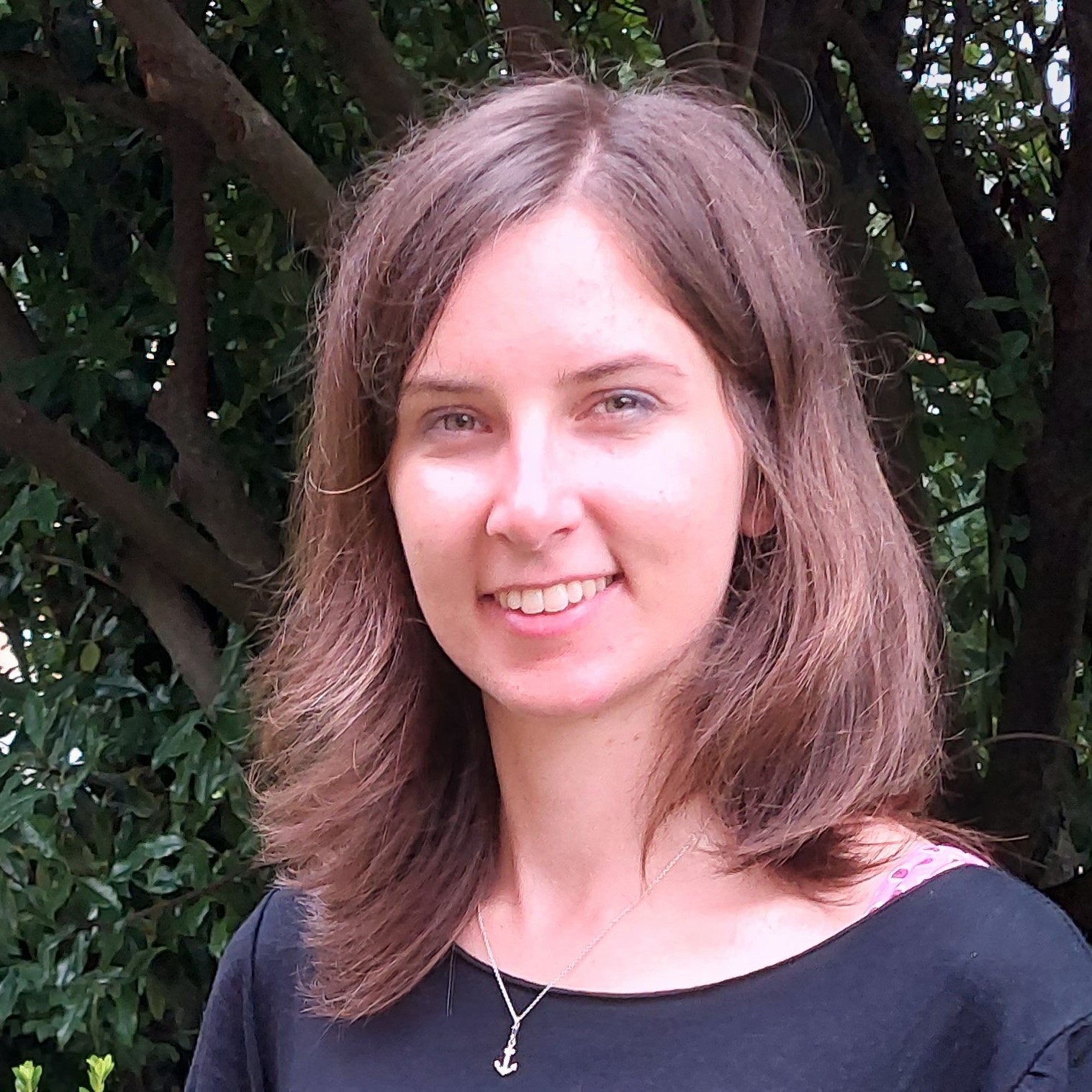}}
\begin{document}
\vspace*{4cm}
\title{A CAUTIONARY TALE: \\ DARK ENERGY IN SINGLE-FIELD, SLOW-ROLL INFLATIONARY MODELS~\footnote{Contribution to the 2022 Cosmology session of the 56th Rencontres de Moriond}}

\author{SVEVA CASTELLO$^1$~\footnote{E-mail: sveva.castello@unige.ch}, STÉPHANE ILIĆ$^{2,3}$, MARTIN KUNZ$^1$}
\address{$^1$D\'epartement de Physique Th\'eorique, Universit\'e de Gen\`eve, 1211 Geneva 4, Switzerland \\
$^2$IRAP, Université de Toulouse, CNRS, CNES, UPS, Toulouse, France \\
$^3$Université PSL, Observatoire de Paris, Sorbonne Université, CNRS, LERMA, F-75014 Paris, France}

\maketitle\abstracts{The current epoch of accelerated cosmic expansion is postulated to be driven by dark energy, which in the standard model takes the form of a cosmological constant with equation of state parameter $w=-1$. We propose an innovative perspective over the nature of dark energy by drawing a parallel with inflation, which we assume to be driven by a single scalar field, the inflaton. The inflaton was not a cosmological constant, as indicated by the fact that inflation ended and by the \textit{Planck} satellite's constraint of $n_s \neq 1$ at 8$\sigma$ confidence. Therefore, it is interesting to verify whether its equation of state parameter was measurably different from -1. We analyze this question for a class of single-field slow-roll inflationary models, where the hierarchy of Hubble slow-roll parameters is truncated at different orders. Based on the latest \textit{Planck} and BICEP2/Keck data, we obtain a 68\% upper bound of $1+w < 0.0014$ for the three-parameter model, which gives the best description to the data. This provides a cautionary tale for drawing conclusions about the nature of today’s dark energy based upon the non-detection of a deviation from $w=-1$ with current and upcoming cosmological surveys.}

\section{Introduction}
The observed accelerated expansion of the Universe is postulated to be driven by a component with negative pressure, dark energy. In the standard model of cosmology, the latter takes the form of a cosmological constant, $\Lambda$, interpreted as a homogeneously-distributed vacuum energy with equation of state parameter $w\equiv \bar{p}/\bar{\rho} = -1$. Available observations constrain its abundance to approximately 68.9\% of the total density, but do not allow to unveil its properties, which remain one of the key issues in modern theoretical physics.  

As suggested by Ilić \textit{et al.}~\cite{Ilic:2010zp} and Castello \textit{et al.}~\cite{Castello:2021nwa}, a precious insight about the nature of dark energy can be achieved by drawing a parallel with a previous epoch of accelerated expansion: inflation. In the simplest description, inflation is driven by a single scalar field, the inflaton, which can be interpreted as a form of dynamical dark energy. However, it cannot be identified with a pure cosmological constant, since it is thought to have rapidly decayed away at the end of inflation. Thus, it is interesting to verify whether a hypothetical observer would have been able to measure the deviation of $w$ from -1 in the inflationary epoch. In the following, we answer this question by characterizing the evolution of $w$ during inflation in light of the latest data about the temperature and polarization anisotropies of the cosmic microwave background (CMB).

\section{Theoretical framework}
\subsection{The Hubble slow-roll parameters}
We assume that inflation lasts for enough e-foldings to ensure that the inflaton dominates the total energy density and adopt the Hamilton-Jacobi formalism \cite{Salopek:1990jq}, where the Hubble parameter $H \equiv \dot{a}/a$ \footnote{Dots and primes indicate derivatives with respect to cosmic time and $\phi$ respectively.} is the reference quantity. Under the standard assumption of slow-roll, we consider the hierarchy of  ``Hubble slow-roll" (HSR) parameters \cite{Liddle:1994dx} $\xi_n$, where each parameter is proportional to the time derivative of the previous one. The first two parameters take the form
\bea
     & \xi_1 (\phi) \equiv \epsilon_H (\phi) = 2 M^2_{Pl} \left(\frac{H'(\phi)}{H(\phi)} \right)^2 \label{epsilonH} \\
    & \xi_2 (\phi) \equiv \eta_H (\phi) = 2 M^2_{Pl} \, \frac{H''(\phi)}{H(\phi)}, \label{etaH}
\eea
with the reduced Planck mass $M_{Pl} \equiv 1/\sqrt{8 \pi}$ and $G = c = \hbar = 1$. 

The first HSR parameter encodes the deviation of $w$ from -1 according to the following exact relation,
\be \label{fundamental_w}
    1+w = \frac{2}{3} \xi_1 \, ,
\ee
and, since $\xi_1 \geq 0$ by definition, this implies that values of $w$ below -1 are nonphysical in our modelling. Moreover, the tensor-to-scalar ratio $r$ and the scalar spectral index $n_s$ can be expressed in terms of the first two HSR parameters at lowest order in slow roll:
\bea 
    & r = 16 \, \xi_1 \label{r} \\
    & n_s - 1 = 2 \xi_2 - 4 \xi_1  \label{n_s}.
\eea
Since the \textit{Planck} satellite has constrained the value of $n_s$ away from 1 at the 8$\sigma$ level \cite{Planck:2018vyg}, Eq.~(\ref{n_s}) implies that either $\xi_1$ or $\xi_2$ must be nonzero. According to Eq.~(\ref{fundamental_w}) and the definition of the HSR parameters, we therefore must require that either $w \neq -1$ or $\frac{dw}{dt} \neq 0$, in both cases yielding a strong observational indication that the inflaton was not a cosmological constant. 

\subsection{Inflationary models}
Our inflationary set-up requires to specify a functional form for $H(\phi)$, which we Taylor-expand around an arbitrary pivot value of the inflaton $\phi_*$ as in Lesgourgues \textit{et al.} \cite{Lesgourgues:2007aa}:
\be \label{Taylor_H}
    H (\phi- \phi_*) = \sum_{i=0}^{n} \hat{H}_i \, (\phi-\phi_*)^n.
\ee
By inserting the series in the definitions of the HSR parameters, the $n$-th order Taylor coefficient can be expressed in terms of the first $n$ parameters evaluated at $\phi_*$ and an additional one, $\xi_0^* (\phi) \equiv \frac{ H^4 (\phi) }{64 M^2_{Pl} \, H'^2(\phi)} \biggr|_{\phi_*}$. 

In our analysis, we truncate the Taylor series at an order varying between one and three, corresponding to models with two, three and four nonzero HSR parameters respectively, labelled HSR\{2\}, HSR\{3\} and HSR\{4\}. We do not consider the zero-th order approximation, since, according to Eq.~(\ref{n_s}), $\xi_1 = \xi_2 = 0$ implies $n_s = 1$ and thus contradicts the aforementioned \textit{Planck} constraint.

\section{Data sets and numerical investigation}
The temperature and polarization anisotropies of the CMB provide a suitable observable to constrain inflationary quantities, since they are believed to mirror the primordial curvature perturbations generated during inflation. We employ the 2018 \textit{Planck} likelihoods \cite{Planck:2019nip}$^,$ \cite{Planck:2018lbu} and the joint BICEP2/Keck-WMAP-\emph{Planck} likelihood \cite{BICEP2:2018kqh} in three combinations:
\begin{itemize}
    \item[(i)] \emph{Planck} low-$\ell$ T/E/B and high-$\ell$ TT/TE/EE (dubbed \emph{P18all} here)
    \item[(ii)] \emph{Planck} low-$\ell$ T/E, high-$\ell$ TT/TE/EE, and low-$\ell$ BICEP2/Keck (\emph{P18$+$BK15})
    \item[(iii)] \emph{Planck} low-$\ell$ T/E, high-$\ell$ TT/TE/EE, lensing and low-$\ell$ BICEP2/Keck (\emph{P18$+$lens$+$BK15}).
\end{itemize}
We perform a Markov Chain Monte Carlo (MCMC) parameter estimation for our inflationary models with the ECLAIR suite of codes \cite{Ilic:2020onu} interfaced with the CLASS Boltzmann solver \cite{Blas:2011rf}. We obtain $n_s$, $r$ and the primordial scalar spectral amplitude $A_s$ as outputs of CLASS, while the MCMC chains for the nonzero $\xi^*$ parameters are converted into allowed evolution histories $w(\phi)$. To achieve a more intuitive interpretation, the constraints on $w$ are then translated to $k$-space via the ``horizon-crossing condition", $k = a H(\phi(a))$, fixing the pivot scale to the standard value in CLASS, $k_* = 0.05$ Mpc$^{-1}$, and considering the range $[k_{\rm min}, k_{\rm max}] = [2 \times 10^{-4}, 0.1]$ Mpc$^{-1}$, which encloses the observable \textit{Planck} interval \cite{Planck:2018jri}.

\section{Results}
\subsection{Parameter constraints} \label{parameter_constraints}
\begin{figure} [h]
\begin{minipage}[t]{0.5\linewidth}
Fig.~\ref{xi_1} shows a strong model-dependence of the MCMC constraints on the parameter $\xi^*_1$, with interesting cosmological implications. In the HSR\{2\} model, where $\xi_2$ is zero, Eq.~(\ref{n_s}) requires a non-trivial value of $\xi_1$ to satisfy the \textit{Planck} constraint of $n_s \neq 1$. This implies the presence of primordial tensor perturbations ($r \neq 0$ from Eq.~(\ref{r})) and yields a detection of $w \neq -1$ according to Eq.~(\ref{fundamental_w}), as is clearly portrayed in Fig.~\ref{w_k_summary}. On the other hand, $\xi_1^*$ is compatible with zero in models with more parameters, such that $r$ is pushed to the trivial value and $w$ to -1. As can be noted from Fig.~\ref{fundamental_w}, the bounds on $w$ get weaker in the HSR\{4\} model with respect to HSR\{3\} and blow up at the boundaries of the chosen $k$-interval, in agreement with the constraints on the observable \textit{Planck} scales \cite{Planck:2018jri}. 
\end{minipage}
\begin{minipage}[t]{0.05\linewidth}
\end{minipage}
\begin{minipage}[t]{0.45\linewidth}
\centerline{\includegraphics[align=t, width=1.0\linewidth]{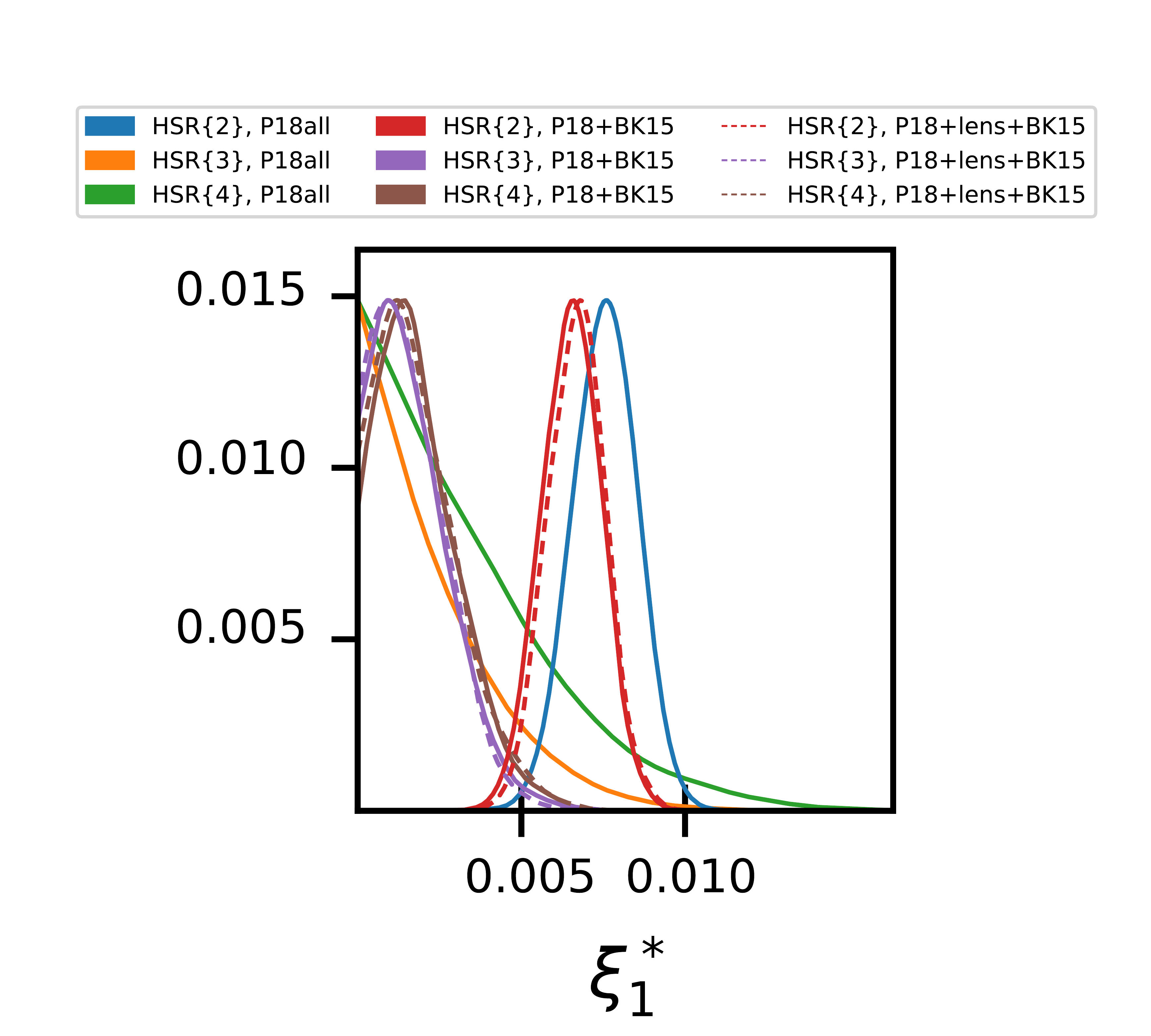}}
\caption{The marginalized constraints on $\xi^*_1$ for all combinations of our tested inflationary models and likelihoods.}
\label{xi_1}
\end{minipage}
\end{figure}
\begin{figure}
    \centering
    \includegraphics[width=0.9\linewidth]{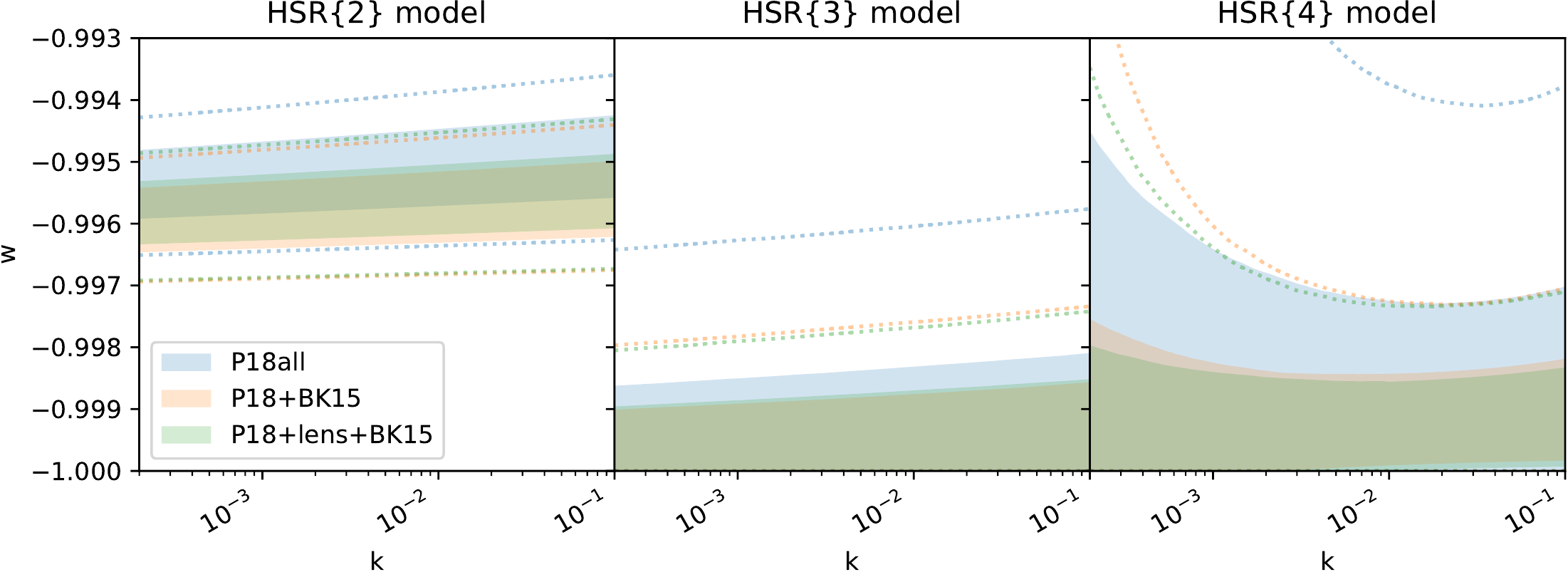}
    \caption{The evolution of $w$ as a function of $k$ at horizon crossing, for each possible pairing of our three inflationary models (left, middle and right panels) and three combinations of likelihoods (in blue, orange and green). The shaded areas correspond to the 68$\%$ confidence intervals and the dotted lines indicate the 95$\%$ ones.}
    \label{w_k_summary}
\end{figure}

\subsection{Model comparison}
We identify the model providing the best description to the data sets through the tools of Bayesian model comparison. Since the HSR\{n\} model is nested into HSR\{n+1\} at $\xi^*_n = 0$, we can compute the Bayes factor $B$ according to the Savage-Dickey Density Ratio \cite{Trotta:2005ar}, i.e.~as the ratio of the prior to the posterior at the nested point, marginalized over the common parameters. We consider two possible choices of flat priors: the ``wide prior" $[-1, 1]$, since we expect the HSR parameters to be smaller than, but generally of order 1, and the ``slow-roll (SR) prior" $[- 0.04, 0.04]$, based on the deviation from a scale-invariant spectrum $|n_s - 1| \approx 0.04$ \cite{Planck:2018vyg}.

The fit to the \textit{P18all} data set yields $B = 11.1$ in favour of HSR\{2\} over HSR\{3\} with the wide prior and gives an undecided outcome with the SR prior. When including the \textit{BK15} data set, however, we obtain a preference for HSR\{3\} in both cases, with $B = 1.3$ for the wide prior and $B = 31$ for the SR prior. The inclusion of \textit{P18lens} strengthens the preference for HSR\{3\} and the comparison with HSR\{4\} indicates that the additional parameter in this model is not constrained by the data.

\section{The cautionary tale}
Based on the analysis with the combined \textit{Planck} and BICEP2/Keck data sets, we conclude that the model with three nonzero parameters, HSR\{3\}, provides the overall best description. This is a non-trivial result, since, as discussed in Sec.~\ref{parameter_constraints}, the choice of model has important cosmological implications in terms of the value of the tensor-to-scalar ratio $r$ and the equation of state parameter of the inflaton $w$.

We obtain a 68\% upper bound of $1+w < 0.0014$ for HSR\{3\}, which is around one order of magnitude below the expected precision of the next generation of cosmological surveys \cite{Euclid:2019clj}. Thus, provided that the current epoch of accelerated expansion can be compared to inflation, our analysis suggests that the lack of an observational detection of $w \neq -1$ in the upcoming decade does not necessarily indicate that today's dark energy is a cosmological constant.

\end{document}